%Paper: hep-th/9206039
%From: GOLDSTONE@irene.mit.edu
%Date: Tue, 9 Jun 1992 13:32:49 -0400 (EDT)

%This is a generic, plain-TeX file requiring no input macros.

\def\pmb#1{\setbox0=\hbox{$#1$}%
\kern-.025em\copy0\kern-\wd0
\kern.05em\copy0\kern-\wd0
\kern-.025em\raise.0433em\box0 }

\def\ick{\eqalignno}

\def\overrightarrow#1{\vbox{\ialign{##\crcr
    \rightarrowfill\crcr\noalign{\kern-1pt\nointerlineskip}
    $\hfil\displaystyle{#1}\hfil$\crcr}}}
\def\overleftarrow#1{\vbox{\ialign{##\crcr
    \leftarrowfill\crcr\noalign{\kern-1pt\nointerlineskip}
    $\hfil\displaystyle{#1}\hfil$\crcr}}}

\def\ick{\eqalignno}
\def\sqr#1#2{{\vcenter{\hrule height.#2pt
   \hbox{\vrule width.#2pt height#1pt \kern#1pt
       \vrule width.#2pt}
         \hrule height.#2pt}}}

\def\_{^{}_}

%\font\brm=cmr10 at 15pt
%\font\cs=cmcsc10
%\vsize=7.5in
%\hsize=5.6in
\tolerance 10000

\baselineskip 12pt plus 1pt minus 1pt
\pageno=0
\rm
\magnification=1200

\centerline {{\bf EXTENDED SUPERCONFORMAL GALILEAN SYMMETRY }}
\vskip 6pt
\centerline{{\bf IN CHERN-SIMONS MATTER SYSTEMS }
\footnote{*}{This
work is supported in part by funds
provided by the U. S. Department of Energy (D.O.E.) under contract
\#DE-AC02-76ER03069, and the World Laboratory.\smallskip}}
\vskip 24pt
\centerline{{M.~Leblanc and G.~Lozano}
\footnote{{\dag}}{On leave from Departamento de Fisica, Universidad
Nac. de la Plata, La Plata, Argentina.\smallskip}}
\vskip 12pt
\centerline{\it Center for Theoretical Physics}
\centerline{\it Laboratory for Nuclear Science}
\centerline{\it and Department of Physics}
\centerline{\it Massachusetts Institute of Technology}
\centerline{\it Cambridge, Massachusetts\ \ 02139\ \ \ U.S.A.}
\vskip 24pt
\centerline{H.~Min}
\vskip 12pt
\centerline{\it Division of Liberal Arts}
\centerline{\it Seoul City University }
\centerline{\it Seoul, 130-743, Korea }
\vskip 1.5in
\centerline{Submitted to: {\it Annals of Physics  \/}}
\vfill
\vskip -12pt
\noindent {CTP\#2106 --- SNUTP 92-37 \hfill June 1992 }
\eject
\baselineskip 24pt plus 2pt minus 2pt
\centerline{\bf ABSTRACT}
\medskip

We study the nonrelativistic limit of the $N=2$ supersymmetric Chern-Simons
matter system. We show that in addition to Galilean invariance the model
admits a set of symmetries generated by fermionic charges, which can be
interpreted as an {\it extended Galilean supersymmetry }.
The system also possesses a hidden conformal invariance and
then the full group of symmetries is the {\it extended
superconformal Galilean} group.
We also show that imposing extended superconformal Galilean symmetry determines
the values of the coupling constants in such a way that their values
in the bosonic sector agree with the values of Jackiw and Pi
for which self-dual equation exist. We finally analyze the
second quantized version of the model and the two-particle sector.
\vfill
\eject
\noindent{\bf I.\quad INTRODUCTION }
\medskip
\nobreak

In the last years much attention has been directed toward the study of matter
systems in (2+1) dimensional space-time minimally coupled to gauge fields
whose dynamics is governed by the Chern-Simons term. The interest in this
subject is not only due to its mathematical richness but also because of its
applications to condensed matter phenomena such as the quantum Hall effect
and high-$T_c$ superconductivity.

An important line of development has followed the papers by
Hong, Kim and Pac [1] and by Jackiw and Weinberg [2]
where vortex solutions that satisfy self-dual
equations were found for relativistic bosons with specific
$\phi^6$-potential [3]. It was
then shown by Lee, Lee, and Weinberg [4] that this specific potential arises
as a consequence of demanding a $N=2$ supersymmetric extension of the
bosonic Abelian Chern-Simons model.

Later Jackiw and Pi [5] introduced
a nonrelativistic model for bosonic fields,
which also supports self-dual solutions when a specific
value for the coupling constant of the Dirac
$\delta$-interaction is chosen.

We explore in the present paper the
supersymmetric extension of the Jackiw-Pi model. In principle, one could
follow different approaches. One of them would be to construct from the
Galilean
algebra in (2+1)-dimensions the graded algebra by either
imposing consistency conditions when fermionic charges are introduced
or by taking the $c\rightarrow\infty$
contraction of the graded Poincar\'e algebra ($c$ is the velocity of light).
Once one has the graded
algebra, one can construct the superfield formalism in order to find
representations of the algebra and write invariant Lagrangians.
This way has
been followed in (3+1)-dimensional theories by Puzalowski [6] as well as by
Azc\'arraga and Ginestar [7] for the
consistency and contraction methods respectively.

We shall instead pursue a different
approach, which is related to the contraction method but
inverts the procedure. Our method consists in taking the nonrelativistic limit
of the relativistic Lagrangian rather than of the
supersymmetric Poincar\'e algebra. Once one has the nonrelativistic
Lagrangian, one can study its symmetries. In particular for the
supersymmetry one can ask whether the set of transformations obtained by taking
the nonrelativistic limit of the supersymmetric Poincar\'e
transformations is still
an invariance of the nonrelativistic Lagrangian. Knowledge of this set of
transformations allows one to construct the fermionic charges and then study
their algebra.

The model enjoys special characteristics.
One of them is the combination of Galilean and local gauge invariances.
Choosing the Abelian Chern-Simons term as the kinetic term for the gauge field
allows one to implement local gauge invariance without introducing
massless particles in the theory, the presence of which would
obstruct Galilean invariance.
Another feature present in our model
is that as we start from a relativistic $N=2$ supersymmetric
Lagrangian we find
that {\it two} fermionic charges (and their Hermitian conjugates) survive
the nonrelativistic limit providing
us then with an {\it extended} Galilean supersymmetry. Although some comments
on extended Galilean supersymmetry have been
made in Refs.~.[6,7],
to our knowledge ours is the first explicit example of a
nonrelativistic Lagrangian
supporting this extended supersymmetry.
Finally, the model admits a hidden SO(2,1) conformal invariance that is not
present in the relativistic model. Thus the full invariance group of the
model is the extended superconformal Galilean group.

The paper is organized as follows. In section II, we introduce the Lagrangian
describing our model by taking the nonrelativistic limit of the $N=2$
supersymmetric Abelian Chern-Simons theory given by Lee, Lee and Weinberg [4].
In section III,  we consider the space-time symmetries of the model.
We start by studying the transformations associated with the Galilean
invariance and we then consider the SO(2,1) conformal symmetry.
In section IV,  we show that for a specific choice of coupling constants
our model possesses an extended supersymmetry.
We construct the charges associated with this symmetry and study its algebra.
We close the
section by writing the self-dual equations of the supersymmetric
model. In section V, after imposing canonical commutation relations for
second quantization, we analyze
the two-particle sector of the model. Final remarks and suggestions for
further developments are left for the concluding section.

\goodbreak
\bigskip
\hangindent=21pt\hangafter=1
\noindent{\bf II. NONRELATIVISTIC LIMIT OF N=2 SUPERSYMMETRIC CHERN-SIMONS
\hfill\break THEORY.}
\medskip
\nobreak

Our starting point is the $N=2$ supersymmetric Chern-Simons theory
described by the action
\footnote{*}{Our conventions for the $\gamma$-matrices are
$\gamma^\mu=(\sigma_3,i\sigma_2,-i\sigma_1)$, $\gamma^\mu\gamma^\nu=
\eta^{\mu\nu}-i\epsilon^{\mu\nu\lambda}\gamma_\lambda$ and the metric
is chosen to be $\eta^{\mu\nu}={\rm diag }(1,-1,-1)$. The letter $x$
denotes the 3-vector $(t,{\bf r})$ and time is frequently omitted in
arguments of quantities taken at equal times for canonical manipulation.
\smallskip} [4]:
$$\ick {
S=\int d^3x \;\;&\Bigl \{ {\kappa\over 4c} \epsilon^{\mu\nu\lambda} A_\mu
F_{\nu\lambda} + |D_\mu\phi|^2 +i{\overline \psi}\gamma^\mu D_\mu\psi\cr
&-({e^2\over \kappa c^2})^2 |\phi |^2\bigl [|\phi|^2-v^2\bigr ]^2
+{e^2\over \kappa c^2}\bigl [3|\phi|^2-v^2\bigr ]{\overline \psi}\psi
\Bigr \} &(2.1a) \cr }
$$
where $\phi$ is a complex spin-0 field carrying two degrees of freedom
and $\psi$ is a two complex components spinor representing spin-1/2
particles also carrying two degrees of freedom.
(Remember that in 2+1 space-time dimensions a Dirac spinor describes a particle
and an antiparticle each with only one spin degree of freedom.) These fields
are minimally coupled to a gauge field (whose dynamics is governed
solely by a topological Chern-Simons term) through the covariant
derivative

$$D_\mu=\partial_\mu +{ie\over c}A_\mu\quad . \eqno (2.1b)$$

In the symmetric phase there are no degrees of freedom associated with
the gauge field and the theory presents the proper counting of degrees
of freedom required by supersymmetry.

The action (2.1a) is invariant under
the following $N=2$ supersymmetric transformations
$$\ick {
\delta A^\mu &= -{e\over \kappa c}({\overline \psi}\gamma^\mu\eta\phi +
{\overline \eta}\gamma^\mu\psi\phi^*)\cr
\delta\phi &={\overline\eta}\psi  \cr
\delta\psi &=-i\gamma^\mu\eta D_\mu\phi + {e^2\eta\over \kappa c^2}
\phi [|\phi|^2 - v^2 ]\quad , &(2.2) \cr }
$$
where $\eta$ is a spinor with two complex components. The
transformations (2.2) are generated by ${\overline \eta}Q
+ {\overline Q}\eta$ where the spinor supercharge is given by:

$$
Q={1\over c}\int d^2r \quad \Bigl \{ \gamma^\mu\gamma^0\psi(D_\mu\phi)^*
-{ie^2\over \kappa c^2} \gamma^0\psi\phi^*[|\phi|^2-v^2] \Bigr \}.
\eqno (2.3)
$$

The nonrelativistic limit of the model can be performed
as follows. We observe first that the quadratic term in the scalar field
defines the boson mass to be $m$,
$$
v^2= {mc^3|\kappa|\over e^2}, \quad\quad (v^2>0)
\eqno(2.4)
$$
and as a consequence of supersymmetry the mass for fermions is also $m$.

In order to simplify the notation, we shall consider only the model with
$\kappa > 0 $. (The theory for negative $\kappa$ can be obtained by
a parity transformation since this is equivalent to change the sign
of $\kappa$.)
The matter part of the Lagrangian density in eq.(2.1a) can be rewritten in the
following way when eq.(2.4) is used
$$\ick {
{\cal L}_{\rm matter}&= {1\over c^2}|(\partial_t +ie A_0)\phi|^2
-|{\bf D}\phi|^2 -m^2c^2|\phi|^2 + {2me^2\over c\kappa }|\phi|^4
-{e^4\over c^4\kappa^2} |\phi|^6\cr
&+{i\over c}{\overline \psi}\gamma^0(\partial_t+ieA_0)\psi
+i{\overline\psi}{\pmb {\gamma} }\cdot{\bf D}\psi -mc {\overline\psi}\psi
+{3e^2\over c^2\kappa} |\phi|^2 {\overline\psi}\psi &(2.5) \cr }
$$
where ${\bf D}=\nabla - {ie\over c}{\bf A}$ is the spatial part of
the covariant derivative $D_\mu = (D_0, {\bf D})$. Similarly
$A^\mu=(A_0,{\bf A})$ and $\gamma^\mu=(\gamma_0,{\pmb {\gamma} })$.

The nonrelativistic limit of the Chern-Simons theory can now be
carried out. Since there are no degrees of freedom associated with the
gauge field in the symmetric phase, we do not modify its action
in the nonrelativistic limit, while for the matter fields we substitute in
eq.(2.5)
$$\ick {
\phi &= {1\over {\sqrt {2m}}} [ e^{-imc^2t}\Phi + e^{imc^2t}{\hat \Phi}^* ]
&(2.6a)\cr
\psi &= {\sqrt c}\;\; [ e^{-imc^2t}\Psi + e^{imc^2t}\sigma_2 {\hat \Psi}^* ]
&(2.6b) \cr}
$$
where $\Phi$ and ${\hat \Phi}$ are the nonrelativistic fields associated
with the particles and antiparticles respectively (similarly for the
fermion field). We eliminate
the second component of the spinor  $\Psi$ and ${\hat\Psi}^*$
by using their equation of motion to leading order in $c$
\footnote{*}{where for any vector ${\bf V}=(V^1, V^2)$,
$V^\pm = V^1 \pm i V^2$.},
$$\ick {
\Psi_2&={-i\over 2mc} D_+\Psi_1 &(2.7a)\cr
{\hat\Psi}_2^*&={-i\over 2mc} D_-{\hat\Psi}_1^* &(2.7b)\cr }
$$
to obtain a nonrelativistic matter Lagrangian density, after dropping terms
that oscillate as
$c\rightarrow\infty$ and terms of ${\cal O}({1\over c^2})$.  In this Lagrangian
particles and antiparticles are independently conserved, and we work in the
zero antiparticle sector by setting ${\hat\Phi}=0$ and
${\hat\Psi}_1=0$ to obtain
$$\ick {
S= \int d^3x \;\;&\bigl\{ {\kappa\over 2c}(\partial_t {\bf A})\times {\bf A}
- A_0[\kappa B+ e(|\Phi|^2 + |\Psi|^2)]\cr
&+i\Phi^*\partial_t\Phi+ i\Psi^*\partial_t\Psi -{1\over 2m}|{\bf D}\Phi|^2
-{1\over 2m}|{\bf D}\Psi|^2 \cr
&+{e\over 2mc} B |\Psi|^2 + \lambda_1 |\Phi|^4 + \lambda_2
|\Phi|^2|\Psi|^2\bigr\} &(2.8) \cr }
$$
where we have dropped the index 1 on spinors. The coupling
constants are given by
$$
\lambda_1= {e^2\over 2mc\kappa}, \quad \quad \lambda_2=3\lambda_1\quad
\eqno(2.9)
$$
and the magnetic field by
$$
B=\nabla\times {\bf A}\quad .
\eqno (2.10)
$$
[In the plane the cross product is ${\bf V}\times {\bf W}
=\epsilon^{ij}V^iW^j$, the curl of a vector is $\nabla\times {\bf V}
=\epsilon^{ij}\partial_iV^j$, the curl of a scalar is
$(\nabla\times S)^i=\epsilon^{ij}\partial_j S$ and we shall introduce
also the notation ${\bf A} \times {\bf {\hat z}}=\epsilon^{ij}A^j$.]

The nonrelativistic limit of the model of eq.(2.1a) led us to a
system of bosons minimally coupled to a gauge field and self-interacting
through an
attractive $\delta$-function potential of strength $\lambda_1$.
Note that fermions couple to the gauge fields not only through the
covariant derivative but also through the Pauli interaction term.
This non-minimal coupling arises in the nonrelativistic limit as a
consequence of the elimination of the second component of the
spinor reminding us of the spin structure of the fermions. Finally,
there is a contact boson-fermion interaction of strength $\lambda_2$.

We close this section by writing the
classical equations of motion which follow from the action (2.8)

$$\ick {
[i(\partial_t &+ ieA_0) + {1\over 2m} D^2 +2\lambda_1 \rho_B
+\lambda_2\rho_F ]\Phi = 0 &(2.11a)\cr
[i(\partial_t &+ ieA_0)+ {1\over 2m} D^2 +\lambda_2 \rho_B
+{e\over 2mc}B ] \Psi = 0  &(2.11b) \cr
B&= -{e\over \kappa} \rho &(2.12a) \cr
{\bf E}\equiv
-\nabla A^0 &-{1\over c}\partial_t {\bf A} = {e\over c\kappa}
\quad {\bf j}\times {\bf {\hat z}} &(2.12b) \cr }
$$
where
$$\ick {
{\bf j} &= {1\over 2mi} [ \Phi^*{\bf D}\Phi - ({\bf D}\Phi)^*\Phi
+\Psi^*{\bf D}\Psi - ({\bf D}\Psi)^*\Psi
+i \nabla \times \rho_F ] &(2.13) \cr }
$$
and
$$\ick {
\rho_B&=|\Phi|^2 , \quad  \rho_F=|\Psi|^2 , \quad
\rho=\rho_B+\rho_F \quad .&(2.14) \cr }
$$

\medskip
\hangindent=22pt\hangafter=1
\noindent{\bf III. SPACE-TIME SYMMETRIES.}
\medskip
\nobreak

In this section we study the space-time symmetries of our system.
First, it is evident that
the model possesses a Galilean invariance containing as generators
the Hamiltonian $H$ (time translation), the total momentum ${\bf P}$
(space translation), the angular momentum $J$ (rotation), and the
Galilean boost generator ${\bf G}$.
In order to close the algebra one has also to
introduce the mass operator which appears as a central charge
({\it ie.,} commutes with all the other operators). Second, the system
possesses a less obvious conformal invariance containing the generators
of the dilation $D$ and of the special conformal transformation
$K$ which together with the Hamiltonian form an $SO(2,1)$
dynamical invariance group.

\medskip
\noindent {\bf III A. Galilean Invariance }
\medskip

The Galilean invariance of the bosonic sector of the model
has already been discussed by Hagen [8] and by Jackiw and Pi [5].
Note that although the
Galilean invariance of the
action without gauge fields is rather obvious, this is not the case
when local gauge invariance is introduced in the theory. The reason why
the Galilean invariance survives the introduction of local gauge invariance
is a consequence of the fact that there is no massless particle associated with
a Chern-Simons gauge field, and therefore any complication related to the
non-relativistic limit of a massless particle is absent in the model.
On the other hand, due to its topological nature,
the Chern-Simons action assure us
that the Chern-Simons action is not only Galilean invariant but invariant
under any coordinate transformations.

We shall
discuss the symmetries and their generators showing that the presence
of fermions adds little complication. We consider
space-time transformations and the corresponding field variations that leave
the action (2.8) invariant and construct the conserved charges
using Noether's theorem.  In order to obtain the gauge
covariant space-time generators, we use the gauge covariant space-time
transformations as given in ref.[9]. They comprise a conventional space-time
transformation generated by a function on space-time, supplemented by a
field dependent gauge transformation.

We start by considering time translation
$$
\delta t=a\quad , \eqno(3.1a)
$$
where $a$ is a constant. The infinitesimal gauge covariant
time-translation on the fields are
$$\ick {
\delta\Phi &=-aD_t\Phi, \quad
\delta \Psi =-aD_t\Psi, \cr
\delta A^{0}&=0 ,\quad
\delta {\bf A} = a {\bf E} \quad , &(3.1b) \cr }
$$
and the conserved charge found using Noether's theorem is the Hamiltonian
$$\ick {
H= \int d^2r \quad &{1\over 2m} \Bigl [|{\bf D}\Phi|^2
+|{\bf D}\Psi|^2 \Bigr ] -{e\over 2mc}B\rho_F \cr
&-\lambda_1 \rho_B^2-\lambda_2\rho_B\rho_F
\quad .&(3.1c) \cr }
$$

The system is also invariant under space translation,
$$
\delta {\bf r}= {\bf a}\quad ,
\eqno (3.2a)
$$
where ${\bf a}$ is a constant 2-vector. The infinitesimal
gauge covariant translations of the fields
$$\ick {
\delta \Phi&=-{\bf a}\cdot{\bf D} \Phi , \quad
\delta \Psi =-{\bf a}\cdot{\bf D} \Psi , \cr
\delta A^{0}&= {\bf a}\cdot {\bf E}, \quad
\delta {\bf A} = {\bf a}\times {\bf {\hat z}} B \quad , &(3.2b) \cr }
$$
lead to the conserved charge
$$\ick {
{\bf P} &= \int d^2r \;\; {\pmb {\cal {P}}} &(3.2c) \cr
{\pmb {\cal {P}}} &= {1\over 2i} [\Phi^*{\bf D} \Phi - ({\bf D}\Phi)^*\Phi
+ \Psi^*{\bf D} \Psi - ({\bf D}\Psi)^*\Psi ]
\quad .&(3.2d) \cr }
$$

The angular momentum is obtained in a similar way by considering
an infinitesimal rotation
$$
\delta {\bf r}= \theta {\bf {\hat z}}\times {\bf r}  \quad ,
\eqno (3.3a)
$$
where $\theta$ is the rotation angle and the infinitesimal gauge
covariant field transformations read
$$\ick {
\delta \Phi &=-\theta\; {\bf r}\times {\bf D} \Phi ,\quad
\delta \Psi =-\theta \; {\bf r}\times {\bf D} \Psi
-{i\over 2}\theta\Psi ,\quad \cr
\delta A^{0}&=\theta \;\; {\bf r}\times {\bf E} ,\quad
\delta {\bf A}=\theta \;\;{\bf r} B\quad  . &(3.3b) \cr }
$$
The angular momentum obtained from the Noether theorem is
$$
J=\int d^2r \;\;\bigl \{ {\bf r}\times {\pmb {\cal {P}}}
+ {1\over 2} \rho_F
\bigr \}\quad . \eqno(3.3c)
$$

The last term in eq.(3.3c) is the spin associated with the fermion
field and originates from the last term in the transformation
for $\Psi$. As the nonrelativistic limit has led us to a 1-component
fermion, the spin is proportional to the fermion number operator
and is then independently conserved [see below].

Under an infinitesimal Galilean boost:
$$
\delta {\bf r}={\bf v} t
\eqno(3.4a)
$$
the fields transform gauge covariantly as
$$\ick {
\delta \Phi &=(im{\bf v}\cdot {\bf r} - t{\bf v}\cdot {\bf D})
\Phi , \quad
\delta \Psi =(im{\bf v}\cdot {\bf r} - t{\bf v}\cdot {\bf D})
\Psi \cr
\delta A^{0}&= t\;{\bf v}\cdot {\bf E} , \quad
\delta {\bf A} = t\; {\bf v}\times {\bf {\hat z}}\;B \quad , &(3.4b) \cr }
$$
and the conserved charge is found to be
$$
{\bf G}=t{\bf P} - m \int d^2r\;\; {\bf r} \rho \quad . \eqno (3.4c)
$$

The Galilean group is completed with the inclusion of
the mass operators
$$\ick {
M_B&=mN_B=m\int d^2r \;\; \rho_B &(3.5a) \cr
M_F&=mN_F=m\int d^2r \;\; \rho_F &(3.5b) \cr }
$$
where $N_B$ and $N_F$ are the boson and the fermion number operator
respectively and $N=N_B+N_F$ is the total number operator.
The conservation of $N_B$ and $N_F$
arises as a consequence of a $U(1)_B\times U(1)_F$ global symmetry (rather than
from a space-time transformation)
$$\ick {
\delta\Phi&=i\alpha\Phi &(3.5c) \cr
\delta\Psi&=i\beta\Psi \quad .&(3.5d) \cr }
$$

Now we turn to the calculation of the algebra satisfied by the above
conserved charges. In order to do that, we solve Gauss's
law by taking the gauge fields as the following
function of the matter fields as in ref.[8,5]
$$
{\bf A}(t,{\bf r}) = {e\over \kappa} \nabla \times \int d^2r'\quad
 G({\bf r}'-{\bf r})  \; \rho (t, {\bf r}')
\eqno (3.6)
$$
where $G({\bf r})$ is the Green's function for the Laplacian in two dimensions
$$
G({\bf r})= {1\over 2\pi} \ln |{\bf r}| \quad .
\eqno (3.7)
$$
Note the ${\bf A}(t, {\bf r})$ is
presented in eq.(3.6) in the Coulomb gauge and we
prescribe that $\nabla\times G({\bf r})|_{{\bf r}=0}=0$.

The Poisson brackets for functions of the matter fields are defined
from the symplectic structure of the Lagrangian at fixed time to be
$$\ick{
\{F,G\}_{P.B.}&=i\int d^2r {\delta F\over \delta\Phi^*({\bf r})}{\delta G\over
\delta \Phi({\bf r})} - {\delta F\over \delta\Phi ({\bf r})}
{\delta G\over\delta
\Phi^*({\bf r})} \cr
&-i\int d^2r {\delta^r F\over \delta\Psi^*({\bf r})}{\delta^l G\over
\delta
\Psi({\bf r})} + {\delta^r F\over \delta\Psi ({\bf r})}{\delta^l G\over\delta
\Psi^*({\bf r})} &(3.8)
\cr }
$$
where the superscripts ``r" and ``l" refer to right and left derivative
and in particular \footnote{*}{We use for the Poisson
bracket $\{F,G\}$ when both functions $F$ and $G$ are
Grassmann functions and $[F,G]$ otherwise because it is suggestive for the
quantum case.}
$$\ick {
[ \Phi ({\bf r}), \Phi^* ({\bf r}') ] &=  -i\delta^2
({\bf r}- {\bf r}') &(3.9a) \cr
\{ \Psi ({\bf r}), \Psi^* ({\bf r}') \}
&=  -i\delta^2 ({\bf r} - {\bf r}') \quad  &(3.9b) \cr }
$$
at fixed time.

Using the Poisson bracket relations of eqs.(3.8-9) the above
conserved charges can be shown to realize the algebra of the Galilean group
$$\ick {
[ P^i, P^j ] &= [ P^i, H] = [ J, H ]= [G^i,G^j]=0 \quad , &(3.10a) \cr
[J, P^i]&= \epsilon^{ij}P^j  &(3.10b) \cr
[J, G^i]&=\epsilon^{ij}G^j  &(3.10c) \cr
[P^i, G^j] &= \delta^{ij} m\;N &(3.10d) \cr
[H, G^i] &=P^i &(3.10e) \cr }
$$

We point out that the only constraint
in the potential required by Galilean invariance is that
$V=V(|\Phi|^2, |\Psi|^2)$ and be a local function of the fields.

\medskip
\noindent {\bf III B. Conformal Invariance }
\medskip

In addition to the Galilean invariance, the model admits a hidden
dynamical $SO(2,1)$ group of conformal transformations.  The role of
conformal invariance in nonrelativistic quantum mechanics was first
studied by Jackiw [10] as well as
by de Alfaro, Fubini and Farlan [11]
and in the framework of Galilean covariant
field theories by Hagen [12] and by Niederer [13].

The presence of the dilation invariance of our model (2.8) can be
easily understood if we ignore for a moment the interaction with
the gauge fields and the fermions. In this case, we are left
with nonrelativistic bosons interacting through a $\lambda\Phi^4$
interaction corresponding to a $\delta$-Dirac potential which is known to be
scale free [14].
This can be seen in the following way: in a nonrelativistic theory we are free
to choose units in such a way that $\hbar$ and $m$ are dimensionless
(and then $[t]$=$[r^2]$). It is then straightforward to check that
$\lambda_1$ is dimensionless.
The interaction with the gauge fields does not change the picture since its
introduction does not require dimensionful coupling constants [5].
On the other hand, looking at the fermionic part of the Lagrangian
we see that it looks like the bosonic counterpart
except for the Pauli interaction term.
However, the Pauli interaction term contains the magnetic field which when
substituted by the Gauss law brings the Pauli interaction in the
same form as the other contact interactions.

Then, under an infinitesimal dilation transformation
$$\ick {
\delta t&=2\alpha t\cr
\delta {\bf r}&=\alpha {\bf r} &(3.11a) \cr }
$$
bosons and fermions transform in the same gauge-covariant way
$$\ick {
\delta \Phi &=-\alpha [ 1 + {\bf r}\cdot {\bf D} + 2tD_t ] \Phi \cr
\delta \Psi &=-\alpha [ 1 + {\bf r}\cdot {\bf D} + 2tD_t ] \Psi &(3.11b)
\cr }
$$
while the transformations for the gauge fields (which we consider
here as independent variables) are
$$\ick {
\delta A^{0}&=\alpha\; {\bf r}\cdot {\bf E} \cr
\delta {\bf A}&=\alpha({\bf r}\times {\bf {\hat z}} \; B + 2t\; {\bf E}) \quad
{}.
&(3.11c)\cr}
$$
It is easy to verify that the Lagrangian (2.8) for arbitrary coupling constants
$\lambda_1$ and $\lambda_2$ is invariant under the
dilation transformations (3.11) and that the conserved charge is
$$
D=Ht - {1\over 2} \int d^2r \quad
{\bf r}\cdot {\pmb {\cal {P}}} \quad .\eqno(3.11d)
$$

As in the purely bosonic model, our system is invariant
under infinitesimal special conformal transformation
$$\ick {
\delta t&= -a\; t^2 \cr
\delta {\bf r}&= -a\; t{\bf r} &(3.12a) \cr}
$$
where $a$ is a constant and the gauge covariant field transformations are
$$\ick {
\delta \Phi &=(at-{i\over 2}mar^2 +at {\bf r}\cdot {\bf D} +at^2D_t )
\Phi \cr
\delta \Psi &=(at-{i\over 2}mar^2 +at {\bf r}\cdot {\bf D} +at^2D_t )
\Psi \cr
\delta A^{0}&=-a\; t{\bf r} \cdot {\bf E}  \cr
\delta {\bf A}&=-a\; t{\bf r}\times {\bf {\hat z}} \; B-a \; t^2{\bf E}
\quad . &(3.12b) \cr }
$$
The conserved charge associated with the special conformal transformation
is given by
$$
K=-t^2H +2tD +{m\over 2}\int d^2r\;\; r^2 \rho\quad .
\eqno (3.12c)
$$

As noted in ref.[5], an energy-momentum tensor can be defined
in such a way that the time independence of $H$ and ${\bf P}$ are assured
by continuity equations of the form

$$\ick {
\partial_tT^{00} +\partial_iT^{i0}&=0 &(3.13a) \cr
\partial_tT^{0i} +\partial_jT^{ji}&=0 &(3.13b) \cr }
$$
where
$$\ick {
T^{00}&={\cal H} &(3.14a) \cr
T^{0i}&={\cal P}^i &(3.14b) \cr
T^{i0}=-{1\over 2m}&\Bigl \{(D_t\Phi)^*D_i\Phi + (D_i\Phi)^*D_t\Phi
+(D_t\Psi)^*D_i\Psi + (D_i\Psi)^*D_t\Psi -{e^2\over \kappa}j^i\rho_F \Bigr \}
\quad \quad &(3.14c)\cr
T^{ij}={1\over 2m} & \Bigl \{ (D_i\Phi)^*(D_j\Phi)+(D_j\Phi)^*(D_i\Phi) +
                    (D_i\Psi)^*(D_j\Psi)+(D_j\Psi)^*(D_i\Psi)\cr
& -  \delta^{ij}[(D_k\Phi)^*(D_k\Phi)+(D_k\Psi)^*(D_k\Psi)]
\Bigr \} \cr
&+ {1\over 4m} (\delta^{ij}\nabla^2-2\partial_i\partial_j)\rho
+\delta^{ij}{\cal H } \quad . &(3.14d) \cr }
$$
Note that $T^{ij}$ is symmetric and has been improved in such a way that
$$\ick {
\delta^{ij}T^{ij} - 2 T^{00} &=0 \quad . &(3.15) \cr }
$$
Formula (3.15) is analogous to the traceless condition of the energy-momentum
tensor for conformal invariance in relativistic theories. The
conservation of the generators $D$ and $K$ is insured by eqs.(3.13-15).

In order to calculate the algebra satisfied by $D$, $K$, and $H$ we use
the Poisson brackets of eqs.(3.8-9), and we consider the gauge field
${\bf A}(t, {\bf r})$ as in eq.(3.6). These charges can be shown to satisfy
the algebra of the conformal SO(2,1) group
$$\ick {
[D,H]&=-H &(3.16a) \cr
[D,K]&=K &(3.16b) \cr
[H,K]&=2D &(3.16c) \cr }
$$
while the algebra of the conformal-Galilean group is closed with the following
additional Poisson brackets
$$\ick {
[K,J]&=[K,G^i]=0 \quad ,[K, P^i]=-G^i &(3.17a) \cr
[D,P^i]&=-{1\over 2} P^i \quad ,[D,J]=0\quad ,
[D,G^i]={1\over 2} G^i \quad .&(3.17b) \cr}
$$
The boson and fermion number generators $N_B$ and $N_F$ have vanishing
Poisson brackets with all generators.

Finally, we note that
the conformal symmetry does not fix completely the potential of the model
but constrains it to be quartic in the fields. The values
of these coupling constants are still arbitrary.

%
%Add the energy momentum tensor here and give it the equation number (3.8)
%comment on traceless, and being the gauged improved tensor.

\medskip
\noindent{\bf IV. EXTENDED GALILEAN SUPERSYMMETRY AND SELF-DUALITY. }
\medskip

Although the idea of supersymmetry is traditionally linked to the
grading of Poincar{\'e}'s group, there has been some development
on $N=1$ supersymmetry in the Galilean framework in 3+1 dimensional
theories [6,7].

As we have already mentioned, the difference between
bosons and fermions are less important
in the nonrelativistic theory,
for instance, with the exception of the Pauli interaction, both particles
have the same kinetic term.  Thus it is not too difficult to imagine
a symmetry exchanging bosons and fermions which behaves as an internal symmetry
though generated by a fermionic charge. Considering the gauge fields as
independent variables, one can check that the following transformation
$$\ick {
\delta_1 \Phi &= {\sqrt {2m}}\quad \eta^*_1\Psi &(4.1a)\cr
\delta_1 \Psi &= -{\sqrt {2m}}\quad \eta_1\Phi &(4.1b)\cr
\delta_1 {\bf A} &= 0 &(4.1c) \cr
\delta_1 A^0 &= {e\over {\sqrt {2m}}\;\;\;c\kappa} (\eta_1\Psi^*\Phi
-\eta^*_1\Psi\Phi^*) \quad , &(4.1d) \cr }
$$
where $\eta_1$ is a complex Grassmann variable, is a symmetry of the
action of eq.(2.8) provided that the following relation holds
$$
2\lambda_1-\lambda_2+{e^2\over 2mc\kappa} = 0\quad .
\eqno(4.2)
$$
The transformation (4.1) can be obtained as the nonrelativistic limit of
eq.(2.2). In a sense, this transformation is of the same kind of
$N=1$ Galilean supersymmetry discussed in ref.[6,7] with
the obvious difference that we are in 2+1 space-time dimensions and
that our gauge fields are not propagating. The last fact provides
us with a simple model incorporating Galilean supersymmetry and local
gauge invariance, avoiding the usual complication of the presence of massless
gauge particles.

We now ask what happens when we consider the next to leading order of the
nonrelativistic limit of eq.(2.2), which is given by
$$\ick {
\delta_2 \Phi &= {i\over {\sqrt {2m}}}\quad\eta^*_2D_+\Psi &(4.3a) \cr
\delta_2 \Psi &= -{i\over {\sqrt {2m}}}\quad\eta_2D_-\Phi  &(4.3b) \cr
\delta_2 A^+ &= {2e\over {\sqrt {2m}}\quad\kappa} \quad
\eta_2\Psi^*\Phi &(4.3c)\cr
\delta_2 A^- &=- {2e\over {\sqrt {2m}}\quad \kappa} \quad
\eta_2^*\Psi\Phi^* &(4.3d)\cr
\delta_2 A^0 &= {i\over (2m)^{3/2}c\kappa}
[ \eta_2(D_+\Psi)^*\Phi + \eta_2^* D_+\Psi \Phi^* ] \quad . &(4.3e) \cr }
$$
The transformation (4.3) is also a symmetry of the action (2.8)
provided that the coupling constants take the values of eq.(2.10). Note
that as in the relativistic case, it is
the second supersymmetry that fixes completely the parameters
of the model, while the transformations (4.1) provide us with a broader class
of Lagrangian according to (4.2).  In particular the value for $\lambda_1$ is
the one for which Jackiw and Pi [5] have found self-dual solution in the
purely bosonic sector.

%The supersymmetric transformations
%(4.1) and (4.3) are given by Poisson bracketing the fields with the charges
%$\eta^*_1 Q_1+ Q_1^*\eta_1 - \eta^*_2Q_2 - Q_2^*\eta_2$ where the
%charges $Q_1$ and $Q_2$ can be obtained by Noether's theorem
Using Noether's theorem, the supersymmetric transformation (4.1) and (4.3)
lead to the charges $Q_1$ and $Q_2$ given by
$$\ick {
Q_1 &= i {\sqrt {2m}}\int d^2r \quad \Phi^* \Psi &(4.4a) \cr
Q_2 &= {1\over {\sqrt {2m}}}\int d^2r \quad
\Phi^* D_+ \Psi \quad . &(4.4b) \cr }
$$

These charges can also be obtained from the nonrelativistic limit of
the supersymmetric spinor supercharge of eq.(2.3). $Q_1$ and $Q_2$ correspond
to leading order of the upper and lower components of the spinor supercharge.

We now study the grading of the conformal-Galilean
group. Solving the constraint of eq.(2.12a) first and
specializing the Hamiltonian of eq.(3.1c) with
the coupling constants given by
eq.(2.10), the Hamiltonian density takes
the simple form
$$\ick {
{\cal H} &= {1\over 2m} [ |{\bf D}\Phi|^2
+|{\bf D}\Psi|^2 ]
-{e^2\over 2mc\kappa } \rho^2
&(4.5) \cr }
$$
and is now supersymmetric. Using the Poisson brackets of eqs.(3.8-9),
the supersymmetry algebra takes the form
$$\ick {
\{ Q_1, Q_1^{*} \} &= -2i M &(4.6a) \cr
\{ Q_2, Q_2^{*} \} &= -iH &(4.6b) \cr
\{ Q_1, Q_2^{*} \} &= -iP_-&(4.6c) \cr
\{ Q_\alpha, Q_\beta \} &= \{ Q^{*}_\alpha , Q^{*}_\beta\} =0 &(4.6d) \cr }
$$
where $Q_1$ and $Q_2$ are given by eqs.(4.4). At this point it is worth to
reproduce eqs.(4.6) by taking the nonrelativistic limit of the algebra
satisfied by the relativistic charge $Q$ of eq.(2.3) [4]
$$\ick {
cP_0&=\pm T + c\{ Q_\pm , Q_\pm^{*} \} &(4.7a) \cr}
$$
where
$$\ick {
Q_\pm &={1\over 2} (1\pm \gamma^0)Q &(4.7b) \cr }
$$
and $T$ is the central charge of the algebra of the relativistic model
$$\ick {
T&= {ev^2\over c}\int d^2x\;\; B \quad .&(4.8) \cr }
$$
The nonrelativistic limit is realized by noticing that
$$\ick {
cP_0 &\rightarrow  Mc^2 + H_{\rm NR} \cr
Q_+&\rightarrow {\sqrt {c}}Q_1 &(4.9)\cr
Q_-&\rightarrow {1\over {\sqrt {c}} } Q_2 \cr }
$$
while for $T$, using Gauss's law and the definition of $v^2$, we obtain
$$
T=-c^2 M \quad .\eqno(4.10)
$$
Thus, we see that the central charge of the relativistic $N=2$ supersymmetric
algebra reduces to the rest energy in the nonrelativistic limit. Therefore
for the upper sign in eq.(4.7a) the central charge adds to the rest energy
to give eq.(4.6a) while for the lower sign the central charge cancels
the rest energy to give eq.(4.6b).

The rest of the Poisson brackets for the graded Galilean
algebra are found to be
$$\ick {
[ P^i, Q_\alpha ] &=[H,Q_\alpha ] = 0, &(4.11a) \cr
[J, Q_1 ] &= {i\over 2} Q_1, \quad
[J, Q_2 ] = -{i\over 2} Q_2 &(4.11b) \cr
[G^+, Q_2] &=0,\quad
[G^-, Q_2] =- Q_1, \quad [ G^i, Q_1 ] = 0
&(4.11c) \cr
[Q_\alpha , N_B ] &=iQ_\alpha ,\quad  [Q_\alpha , N_F]=-i Q_\alpha
\quad . &(4.11d) \cr }
$$
{}From the behavior under the rotations and Galilean boost
we see that in fact $Q_1$ transforms as the upper component of a
spinor and $Q_2$ as the lower component.

We now add the conformal generators.
First we note that the crossing between the fermionic charge $Q_1$ and the
conformal algebra as well as $D$ with $Q_2$ does not generate new charges
$$\ick {
[D, Q_1 ] &= [ K, Q_1]= 0, \quad
[D,Q_2]=-{1\over 2} Q_2 \quad  &(4.12) \cr }
$$
however the crossing between the charge $Q_2$ and the special conformal
generator $K$ produces a new generator
$$\ick {
[K, Q_2] &= -i F\quad  &(4.13a) \cr
F&=-itQ_2 - {\sqrt {{m\over 2}}} \int d^2r \;\; {\bf r}^+ \Phi^{*}\Psi
\quad ,&(4.13b) \cr }
$$
which is needed to close the superconformal Galilean algebra and can be
obtained as the charge of the following symmetry transformation
$$\ick {
\delta\Phi&=\xi\bigl ( {t\over {\sqrt {2m}}} D_+ -i {\sqrt {{m\over 2}}}
\quad {\bf r^+} \bigr ) \Psi     &(4.14a) \cr
\delta \Psi&
=\xi^*\bigl (
-{t\over {\sqrt {2m}}} D_- +i {\sqrt {{m\over 2}}}
\quad {\bf r}^-\Phi(y) \quad .   &(4.14b) \cr }
$$

The behavior of this new operator under the Galilean group is given
by the following Poisson brackets
$$\ick {
[F,H]&=iQ_2, \quad
[F, P^+] = 0, \quad
[F,P^-]= -i Q_1, \quad
[F, G^i]=0, \quad  [F,J]={i\over 2} F \quad \quad &(4.15a) \cr }
$$
while the crossing with the conformal generators gives
$$\ick {
[D,F]&={1\over 2} F, \quad [F,K]=0\quad . &(4.15b)\cr }
$$
The Poisson brackets with the fermionic charges are found to be
$$\ick {
\{F, Q_\alpha\} &= 0, \quad
\{F,F^{*}\}= -iK \quad \cr
\{F,Q_1^{*}\}&= - G^+, \quad
\{F,Q_2^{*}\}= {i\over 2}[ 2iD-J+N_B-{1\over 2}N_F ] \quad
&(4.16)\cr
[F,N_B] &=iF, \quad   [F,N_F]= -iF \quad . \cr }
$$

Note that although $N_B$ and $N_F$ do not have vanishing Poisson brackets
with all operators,
$N$ does and therefore $N$ is a central charge.

The 16 generators ${\bf P},J,{\bf G},H,D,K,Q_1,Q_2,Q^{*}_1,Q^{*}_2,F,F^{*},
N_B,N_F$ generate the extended superconformal Galilean algebra of our model.

We close this section by exploring the relation between $N=2$
supersymmetry and self-duality.
We have already shown that the $N=2$ supersymmetry fixes the value of the
coupling constant $\lambda_1$ to be the one for which self-dual solutions
exist in Jackiw-Pi model [5]. Now we would like to investigate how the
presence of the fermions modify the self-dual equations. In order to do this,
we write the Hamiltonian in its self-dual form.
We first collect the identities
$$\ick {
|{\bf D}\Phi|^2&=|D_\pm \Phi |^2\pm m\nabla\times {\bf j}_B
\pm {e\over c} B\rho_B &(4.17a)\cr
|{\bf D}\Psi|^2&=|D_\pm \Psi |^2\pm m\nabla \times {\bf j}_F
\pm {e\over c} B\rho_F \pm {1\over 2} \nabla^2 \rho_F &(4.17b)\cr }
$$
with
$$\ick {
{\bf j}_B&= {1\over 2mi} [ \Phi^*{\bf D}\Phi - ({\bf D}\Phi)^*\Phi ]
&(4.17c) \cr
{\bf j}_F&= {1\over 2mi} [ \Psi^*{\bf D}\Psi -
({\bf D}\Psi)^*\Psi + i \nabla \times \rho_F ] \quad .&(4.17d) \cr }
$$

Using eqs.(4.17) and Gauss's law,
the energy density ${\cal H}$ can be written as
$$\ick {
{\cal H} &= {1\over 2m} [ |D_\pm\Phi|^2 + |D_\pm\Psi|^2 ] \pm {1\over 2}
\nabla\times [{\bf j}_B+{\bf j}_F ] \pm {1\over 4m} \nabla^2 \rho_F \cr
&-[\lambda_1\pm {e^2\over 2mc\kappa}]\rho_B^2
-[\lambda_2\pm {e^2\over mc\kappa}-{e^2\over 2mc\kappa}]\rho_B
\rho_F\quad . &(4.18) \cr}
$$
Consequently, with $\lambda_1=\mp {e^2\over 2mc\kappa}$,
$\lambda_2=(1\mp 2){e^2\over 2mc\kappa}$ and for well behaved fields for
which the integral of $\nabla \times [{\bf j}_B + {\bf j}_F]$ and
$\nabla^2 \rho_F$
vanishes, the Hamiltonian is
$$\ick {
H&=\int d^2r \quad {1\over 2m} [ |D_\pm\Phi|^2 + |D_\pm\Psi|^2 ]\quad
.&(4.19)\cr }
$$
$H$ reaches its minimum value, zero, when
$$\ick {
D_1\Phi&=\mp iD_2\Phi   &(4.20a) \cr
D_1\Psi&=\mp iD_2\Psi \quad . &(4.20b) \cr }
$$
Together with the Gauss law (2.12a) these two equations compose the
{\it super self-duality equation}. When $\Psi$ is set to zero, the above set
of equations reduces to the ordinary self-duality equations in [5].

Solutions for the minimum energy exist only for the lower sign in eqs.(4.20).
This can be seen by the following argument. Decompose the fields
$$
\Phi=e^{i{e\over c}\omega_B}\rho_B^{1/2}\quad {\rm and } \;\;
\Psi=\eta e^{i{e\over c}\omega_f}\rho_F^{1/2}\eqno(4.21)
$$
and substitute in eq.(4.20) to get
$$\ick {
{\bf A}&=\nabla \omega_B\pm{c\over 2e}\nabla\times
\ln \rho_B &(4.22a) \cr
{\bf A}&=\nabla\omega_F\pm{c\over 2e}\nabla\times
\ln \rho_F \quad . &(4.22b) \cr }
$$
Since eqs.(4.22a-b) must be consistent the fermionic
and bosonic densities are proportional.
Substituting eqs.(4.22) in eq.(2.12a), we find
$$\ick {
\nabla^2 \ln \rho &= \pm 2 {e^2\over c\kappa} \rho \quad ; \; \kappa >0
&(4.23) \cr }
$$
Solutions for eq.(4.23) exist only when the constant in front of the
right side of eq.(4.23) is negative. Since we have taken
$\kappa > 0$ solutions exist only for the lower sign and the energy takes the
form
$$\ick {
H&=\int d^2r \quad {1\over 2m} [ |D_-\Phi|^2 + |D_-\Psi|^2 ] \quad .&(4.24)\cr
}
$$
Therefore, the presence of fermions does not modify the bosonic self-dual
equation, a situation which is familiar in several relativistic models [15].

\medskip
\bigskip
\hangindent=22pt\hangafter=1
\noindent{\bf V. SECOND QUANTIZATION. }
\medskip
\nobreak

In this section, we present the second quantized version of the model.
We consider bosonic quantum field operators $\Phi$,
fermionic quantum field operators
$\Psi $ and their hermitian conjugate
$\Phi^{\dag}$, $\Psi^{\dag}$ satisfying equal-time
commutation and anticommutation relations

$$\ick {
[ \Phi ({\bf r}), \Phi^{\dag} ({\bf r}') ] &=  \delta^2({\bf r}-{\bf r}')
&(5.1a) \cr
\{ \Psi ({\bf r}), \Psi^{\dag} ({\bf r}') \}
&=  \delta^2 ({\bf r}-{\bf r}') \quad , &(5.1b) \cr }
$$
and use the gauge fields as
the function of the matter fields of eq.(3.6).

In the following, we shall consider the Hamiltonian
$$\ick {
H&=\int d^2r \;\;\; {\cal H} &(5.2a) \cr
{\cal H} &= {1\over 2m} [ ({\bf D}\Phi)^{\dag}\cdot{\bf D}\Phi
+({\bf D}\Psi)^{\dag}\cdot {\bf D}\Psi ] - {e\over 2mc } :B\rho_F : \cr
&-\lambda_1 :\rho_B^2: - \lambda_2\rho_B\rho_F
&(5.2b) \cr }
$$
where $\lambda_1\;\; , \lambda_2$ are kept arbitrary
for the moment. Note that although ${\bf D}\Phi=
(\nabla - {ie\over c}{\bf A}) \Phi$ is normal ordered,
the term
$({\bf D}\Phi)^{\dag}\cdot({\bf D}\Phi)$ is not (similarly for the fermions).

The quantum field equations of motion follow from the commutation and
anticommutation relations (5.1)
$$\ick {
i\partial_t\Phi (x) &= [ \Phi (x), H ] \cr
&= \Bigl [-{1\over 2m} D^2  + e A_0 - 2\lambda_1\rho_B
-\lambda_2 \rho_F \Bigr ] \Phi (x) \cr
&+ {e^4\over 2mc^2\kappa^2} \int d^2r' \;\; \nabla G({\bf r}'-{\bf r})
\cdot\nabla G({\bf r}'-{\bf r})
\rho(t, {\bf r}') \Phi (x) &(5.3a) \cr
i\partial_t\Psi (x) &= [ \Psi (x), H ] \cr
&= \Bigl [-{1\over 2m} D^2 + e A_0 +
({e^2\over 2mc\kappa }-\lambda_2 ) \rho_B \Bigr ] \Psi (x) \cr
&+ {e^4\over 2mc^2\kappa^2} \int d^2r' \;\; \nabla G({\bf r}'-{\bf r})
\cdot\nabla G({\bf r}'-{\bf r})
\rho(t, {\bf r}') \Psi (x) &(5.3b) \cr }
$$
where the scalar gauge potential in eqs.(5.3) given by
$$
A_0(t, {\bf r})=
-{e\over c\kappa}\int d^2r' \;\; \{G({\bf r}'-{\bf r})
\nabla\times {\bf j}(t, {\bf r}')\}
\eqno(5.4)
$$
solves eq.(2.12b) together with ${\bf A}(t, {\bf r})$
given by eq.(3.6). The current
${\bf j}$ is given as in eq.(2.13). The last term in eqs.(5.3) is not present
in the classical equation of motion and arises here
because of reordering of operators.

%Note that the change in the derivatives does not change $P^i, D, K,
%G^i$ because $\int d^2x\;\; A^i\rho = 0$ and
%$\int \;\; x^iA^i \rho = 0$. However,
%the angular momentum operator picks up an extra term
%$$
%J=J_{\rm old} + {e^2\over 4\pi c\kappa } N^2 \;\; .
%\eqno(5.10)
%$$

%The inclusion of the last term in eq.(5.10) introduce little modification
%in the algebra since $N$ is a central charge.

Now we study the Schr\"odinger equation for the two-particle sector.
We assume the existence of a zero particle state
$|\Omega >$ which is annihilated by

$$\ick {
\Phi\; |\Omega > &= 0 = <\Omega | \Phi^{\dag} &(5.5a)\cr
\Psi\; |\Omega > &= 0 = <\Omega | \Psi^{\dag} &(5.5b)\cr}
$$
and also by
$$\ick {
N_B|\Omega > &= N_F|\Omega >= 0 \quad {\rm and }\quad H |\Omega>=0 &(5.5c)\cr }
$$

We denote a state of energy $E$ with $N_B$ bosons and $N_F$ fermions by
$|E,N_B,N_F>$ and we write the orthonormalized states as
$$\ick {
|E,2,0>&={1\over {\sqrt 2}}\int d^2r_1\; d^2r_2\quad u_B({\bf r}_1,{\bf r}_2)
\Phi^{\dag}({\bf r}_1)\Phi^{\dag}({\bf r}_2)\;|\Omega > &(5.6a)\cr
|E,0,2>&={1\over {\sqrt 2}}\int d^2r_1\; d^2r_2\quad u_F({\bf r}_1,{\bf r}_2)
\Psi^{\dag}({\bf r}_1)\Psi^{\dag}({\bf r}_2)\;|\Omega > &(5.6b)\cr
|E,1,1>&= {1\over {\sqrt {2}}} |E,1,1>_S + {1\over {\sqrt {2}}} |E,1,1>_A \cr
&={1\over 2}\int d^2r_1\; d^2r_2\; \Bigl \{ u_S({\bf r}_1,{\bf r}_2)
[\Phi^{\dag}({\bf r}_1)\Psi^{\dag}({\bf r}_2)+\Phi^{\dag}({\bf r}_2)
\Psi^{\dag}({\bf r}_1)]
\;|\Omega > \cr
& \quad \quad \quad \quad + u_A({\bf r}_1,{\bf r}_2)
[\Phi^{\dag}({\bf r}_1)\Psi^{\dag}({\bf r}_2)-\Phi^{\dag}({\bf r}_2)
\Psi^{\dag}({\bf r}_1)]
\;|\Omega > \Bigr \} &(5.6c)\cr }
$$
where we have divided the one fermion-one boson state in a symmetric
and anti-symmetric state for convenience. The wave functions
$u_B$, $u_F$, $u_S$ and $u_A$ are normalized to unity.

It is also convenient to collect the wave functions in a vector
$$
U=\pmatrix { u_B \cr
             u_S \cr
             u_A \cr
             u_F \cr }
\quad . \eqno (5.7)$$
The Schr\"odinger equation for the wave functions
can be easily found using the equations of motion (5.3) for arbitrary
$\lambda_1 , \lambda_2$ and the
commutation and anticommutation relations (5.1) to be

$$\ick {
E U &= H U &(5.8) \cr
&=\Bigl \{ -{1\over 2m}\Bigl ( {\cal D}^2_1 + {\cal D}^2_2 \Bigr ){\cal I}
-\delta ({\bf r}_1 - {\bf r}_2 )
\pmatrix { 2\lambda_1 & 0 & 0 & 0 \cr
           0 & \lambda_2-{e^2\over 2mc\kappa }  & 0 & 0 \cr
           0 & 0 & 0 & 0 \cr
           0 & 0 & 0 & 0 \cr }  \Bigr \} U \cr }
$$
where ${\cal I}$ is the identity matrix and
$$\ick {
{\bf {\cal D}}_1
& = \nabla_1 - {ie^2\over c\kappa}\nabla_1\times
G({\bf r}_1-{\bf r}_2) \cr
{\bf {\cal D}}_2 & = \nabla_2 - {ie^2\over c\kappa}\nabla_2\times
G({\bf r}_2-{\bf r}_1)\quad . &(5.9)\cr}
$$

The contact interaction for the symmetric wave functions $u_B$ and $u_S$
present in their Schr\"odinger equation differs in general, however, when
the first supersymmetry is imposed the relation (4.2) implies that
$u_B$ and $u_S$ satisfy the same differential equation both with strength
$2\lambda_1$. The second supersymmetry fixes the value of the strength
of the contact interaction to be ${e^2\over mc\kappa }$.

We now study the action of the supersymmetric charges on the wave functions.
The easiest way to do this is to give the matrix form of $Q_1$,

$$Q_1=2i{\sqrt {m}}
\pmatrix { 0 & 1 & 0 & 0 \cr
           0 & 0 & 0 & 0 \cr
           0 & 0 & 0 & 1 \cr
           0 & 0 & 0 & 0 \cr }
\eqno (5.10)$$
and for $Q_2$,
$$Q_2={1\over 2 {\sqrt {m} } }
\pmatrix { 0 & {\cal D}^+_1+{\cal D}^+_2 & -{\cal D}^+_1+{\cal D}^+_2 & 0 \cr
           0 & 0 & 0 & {\cal D}^+_1 - {\cal D}^+_2 \cr
           0 & 0 & 0 & {\cal D}^+_1 + {\cal D}^+_2 \cr
           0 & 0 & 0 & 0 \cr }
\eqno (5.11)$$
where ${\bf {\cal D}}^+_1$ and ${\bf {\cal D}}^+_2$
can be deduced from the expressions (5.9).
It is easy to verify that $Q_1$ and $Q_2$ in eqs.(5.10-11) satisfy the extended
supersymmetry of eqs.(4.6) with $M=2m{\cal I}$ and $H$ as in eq.(5.8)
with $\lambda_1$ and $\lambda_2$ given by eq.(2.10).

\vfill
\eject
\noindent {{\bf VII. DISCUSSION. }}

The goal of the present investigation was to explore the supersymmetric
extension of Jackiw-Pi model.  This was done by taking the nonrelativistic
limit of the $N=2$ supersymmetric model of Lee, Lee, Weinberg. We have
shown that the resulting theory possesses two independent fermionic
symmetries which together with the Galilean symmetry generate the extended
Galilean supersymmetry. In this context, we have shown that two usual results
that are familiar for relativistic models also hold in the framework of
extended Galilean supersymmetry
{\it i.e.,} the fact that the extended supersymmetry
fixes the value of the bosonic potential to be the one which admits self-dual
equations [16] and the fact that even when the Euler-Lagrange equations are
modified by the presence of fermions, the self-dual equations for bosons are
not [15].
The structure of the algebra is also enriched by the presence of an additional
conformal symmetry. We have shown that its presence enlarges the Galilean
supersymmetry algebra to be a superconformal Galilean algebra. The discussion
of this point has been treated classically. It is known that in several related
models that this conformal symmetry is broken quantum mechanically and one can
wonder if this breakdown of conformal symmetry occurs in our model [14,17].

Apart from a careful analysis of the breakdown of the conformal symmetry,
there are several directions to be explored. One of them is to construct
the adequate superspace formalism in (2+1)-dimensions for extended
Galilean supersymmetry and to build our system in this framework. The
superspace formalism, although less direct than our approach, could be
useful for further generalization of our model.
Other interesting problems are the study of the nonrelativistic limit
of the Lee, Lee, Weinberg model but in the symmetry breaking sector or
of the $N=2$ supersymmetric Maxwell-Chern-Simons model which has also a
$(F_{\mu\nu})^2$
term [18]. In both cases, there are massive particles associated
to the gauge fields and it would be interesting to know how their presence
would modify the supersymmetry algebra.

\vfill
\eject
\centerline {ACKNOWLEDGEMENTS. }
We acknowledge Professor
Roman Jackiw for suggesting the problem, and for continuous
and stimulating discussions. We also thank D. Bazeia, R. Brooks,
M. Crescimanno, D.Z. Freedman, and C.K. Lee for discussions.
M.L. thanks N.S.E.R.C. for financial support. G.L.
thanks A. Zichichi and the
World Laboratory for financial support. H.M. is supported
by KOSEF and the Ministry of Education of Korea. M.L. and G.L. thank
J. Negele and the CTP for hospitality. H.M. expresses its gratitude
to the department of physics, Boston University, where this work
has been initiated.

\vfill
\eject
\centerline{\bf REFERENCES }
\medskip
\item{1.}J. Hong, Y. Kim and P. Y. Pac, {\it Phys. Rev. Lett.\/} {\bf 64}
(1990) 2230.
\medskip
\item{2.}R. Jackiw and E. J. Weinberg, {\it Phys. Rev. Lett.\/} {\bf 64} (1990)
2234.
\medskip
\item{3.}For a review on the subject see R. Jackiw and S-Y. Pi,
{\it ``Self-Dual Chern-Simons solitons"}, MIT-CTP-2000.
\medskip
\item{4.}C. Lee, K. Lee and E.J. Weinberg, {\it Phys. Lett.\/} {\bf 253B}
(1990) 105.
\medskip
\item{5.}R. Jackiw and S-Y. Pi, {\it Phys. Rev. D\/} {\bf 42} (1990)3500;
{\it Phys. Rev. Lett.\/} {\bf 64 } (1990)2969.
\medskip
\item{6.}R. Puzalowski, {\it Acta Phys. Austriaca\/} {\bf 50}, (1978)45.
\medskip
\item{7.}J.A. de Azc\'arraga and D. Ginestar, {\it J. Math. Phys.\/}
{\bf 32} (1991)12 and references therein.
\medskip
\item{8.}C.R. Hagen, {\it Phys. Rev.\/} {\bf 31} (1985)848.
\medskip
\item{9.}R. Jackiw, {\it Physics Rev. Lett. \/} {\bf 41 } (1978)1635.
\medskip
\item{10.}R. Jackiw, {\it Physics Today \/} {\bf 25 } (1972)23.
\medskip
\item{11.}V. de Alfaro, S. Fubini and G. Furlan, {\it Nuovo
Cimento \/} {\bf A 34} (1976)569.
\medskip
\item{12.}C.R. Hagen {\it Phys. Rev.\/} {\bf D5} (1972)377.
\medskip
\item{13.}U. Niederer, {\it Helv. Phys. Acta \/} {\bf 47} (1974)119.
\medskip
\item{14.}R. Jackiw, {\it ``M.A.B. B\'eg Memorial Volume" },
(A. Ali, P. Hoodbhoy, Eds.) World Scientific, Singapore, 1991; and
MIT-CTP-1937.
\medskip
\item{15.}P. DiVecchia and S. Ferrara, {\it Nucl. Phys.\/} {\bf B130} (1977)
93.
\medskip
\item{16.}Z. Hlousek and  D. Spector, {\it Nucl. Phys.\/} {\bf B370} (1992)
143.
\medskip
\item{17.}O. Bergman, MIT-CTP-2045, (1991).
\medskip
\item{18.}C. Lee, K. Lee and H. Min, {\it Phys. Lett. \/} {\bf 252B} (1990)
79.
\medskip

\end